\documentstyle[multicol,aps,psfig]{revtex}
\begin{document}
\draft
\preprint{LBNL}

\title{Multiple Scattering, Parton Energy Loss and Modified Fragmentation 
Functions in Deeply Inelastic $eA$ Scattering }
\author{Xiaofeng Guo$^1$ and Xin-Nian Wang$^2$}
\address{$^1$ Department of Physics and Astronomy, University of Kentucky,
Lexington, Kentucky KY 40506}
\address{$^2$ Nuclear Science Division, Mailstop 70-319,
Lawrence Berkeley National Laboratory, Berkeley, CA 94720}

\date{May 4, 2000}

\maketitle

\begin{abstract}
Modified quark fragmentation functions in deeply inelastic $eA$ collisions 
and their QCD evolution equations are derived for the first time in the 
framework of multiple parton scattering. Induced radiation gives rise to 
additional terms in the evolution equations and thus softens the modified 
quark fragmentation functions. The results in the 
next-leading-twist depend on both diagonal and off-diagonal 
twist-four parton distributions and the combination of which clearly 
manifests the LPM interference pattern. The predicted modification 
depends quadratically on the nuclear size ($A^{2/3}$). 
Generalization to the case of hot QCD medium is also discussed.
\end{abstract}


\begin{multicols}{2}

The propagation of an energetic parton and its induced energy loss has been 
proposed as a probe of the properties of dense matter formed in high-energy 
nuclear collisions \cite{qn1,qn2}. Recent theoretical studies 
\cite{GW1,BDMPS,BGZ,GLV} show that a fast parton will lose a significant 
amount of energy via induced radiation when it propagates through a hot partonic 
matter. One cannot directly measure the energy 
loss of partons because they are not final experimentally observed 
particles. However, parton energy loss does lead to modification of
the final particle spectra. Therefore, one can only study the parton energy loss 
indirectly by measuring the modification of the parton fragmentation functions
in semi-inclusive processes 
like $eA$ or $\gamma$-jet events in $AA$ collisions \cite{wh97} or
the inclusive spectra at large transverse momentum \cite{qn2,wang98}.

In this Letter, we report our first study and derivation of the QCD 
evolution equations for the medium-modified fragmentation functions in the
simplest case of deeply inelastic $eA$ scattering (DIS). The induced gluon
radiation due to multiple parton scattering gives rise to additional terms
in the modified QCD evolution equations that soften the modified fragmentation
functions. Utilizing the generalized factorization of higher-twist (HT) parton 
distributions \cite{LQS1}, we show that these additional HT terms 
depend on both the diagonal and off-diagonal twist-four
parton distributions, the combination of which clearly manifests the
Landau-Migdal-Pomeranchuk (LPM) interference pattern. Using estimates of these twist-four parton matrix
elements from other processes such as the $p_T$ broadening of
Drell-Yan dilepton in $pA$ collisions, we predict the modification of the
effective quark fragmentation functions and their dependence on the parton
energy and nuclear size. We also estimate the quark energy loss defined
as the total energy carried by gluons from induced radiation.

We consider the following semi-inclusive process in the deeply inelastic 
$eA$ scattering, $e(L_1) + A(p) \longrightarrow e(L_2) + h (\ell_h) +X$,
where $L_1$ and $L_2$ are the four momenta of the incoming and the 
outgoing leptons, $\ell_h$ is the observed hadron momentum, $p$ and 
$q=L_2-L_1$ denoted as $p = [p^+,0,{\bf 0}_\perp] \label{eq:frame}$,
$q = [-Q^2/2q^-, q^-, {\bf 0}_\perp]$,
are the momentum per nucleon in the nucleus with the atomic 
number $A$ and the momentum transfer, respectively. The differential 
cross section for the semi-inclusive process can be expressed as
\begin{equation}
E_{L_2}E_{\ell_h}\frac{d\sigma_{\rm DIS}^h}{d^3L_2d^3\ell_h}
=\frac{\alpha^2_{\rm EM}}{2\pi s}\frac{1}{Q^4} L_{\mu\nu}
E_{\ell_h}\frac{dW^{\mu\nu}}{d^3\ell_h}
\label{sigma-dis}
\end{equation}
where $s=(p+L_1)^2$ and $\alpha_{\rm EM}$ is the electromagnetic (EM) 
coupling constant. The leptonic tensor is given by 
$L_{\mu\nu}=1/2\, {\rm Tr}(\gamma \cdot L_1 \gamma_{\mu}
\gamma \cdot L_2 \gamma_{\nu})$
while the semi-inclusive hadronic tensor is defined as,
\begin{eqnarray}
E_{\ell_h}\frac{dW_{\mu\nu}}{d^3\ell_h}&=&
\frac{1}{2}\sum_X \langle A|J_\mu(0)|X,h\rangle 
\langle X,h| J_\nu(0)|A\rangle \nonumber \\
&\times &2\pi \delta^4(q+p-p_X-\ell_h)
\end{eqnarray}
where $\sum_X$ runs over all possible final states and 
$J_\mu=\sum_q e_q \bar{\psi}_q \gamma_\mu\psi_q$ is the
hadronic EM current.

In the parton model with collinear factorization approximation and
to the leading-twist (LT) the semi-inclusive cross section factorizes 
into a product of parton distributions, parton fragmentation functions 
and the partonic cross section. Therefore, to the leading order 
in  $\alpha_{\rm s}$,
\begin{eqnarray}
& &\frac{dW^S_{\mu\nu}}{dz_h}
= \sum_q e_q^2 \int dx f_q^A(x,\mu_I^2) H^{(0)}_{\mu\nu}(x,p,q) 
D_{q\rightarrow h}(z_h,\mu^2)\, \nonumber \\
& &H^{(0)}_{\mu\nu}(x,p,q) = \frac{1}{2}\, 
{\rm Tr}(\gamma \cdot p \gamma_{\mu} \gamma \cdot(q+xp) \gamma_{\nu})
\, \frac{2\pi}{2p\cdot q} \delta(x-x_B) \, , \nonumber\\
& &\label{eq:s-sum}
\end{eqnarray}
where the momentum fraction carried by the hadron is defined as 
$z_h=\ell_h^-/q^-$ and $x_B=Q^2/2p^+q^-$ is the Bjorken variable.
$\mu_I^2$ and $\mu^2$ are the factorization scales for the initial
quark distributions $f_q^A(x,\mu_I^2)$ in a nucleus and the fragmentation 
functions $D_{q\rightarrow h}(z_h,\mu^2)$, respectively.
Including all leading log radiative
corrections, the renormalized quark fragmentation function
$D_{q\rightarrow h}(z_h,\mu^2)$ satisfies the QCD evolution
equation \cite{Fields}.

\begin{figure}
\centerline{\psfig{figure=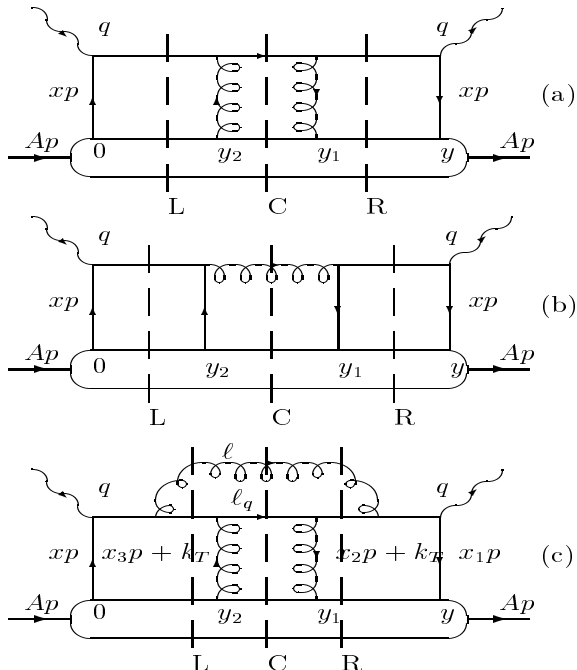,width=3.0in,height=3.5in}}
\caption{Diagrams for rescattering with gluons (a) and quarks (b) without and with
gluon radiation (c) in deeply inelastic $eA$ scattering. Possible cuts
are shown by the dashed lines}
\label{fig1}
\end{figure}

In this paper we will consider contributions of quark rescattering 
with partons from another nucleon inside the nucleus. Such contributions
are proportional to the nuclear size $A^{1/3}$ \cite{LQS1}. For large
enough $A$, we can neglect other $A$-independent HT effects.
For large $Q^2$ in DIS, it may suffice to only consider one rescattering.
The contributions of one rescattering can be treated as HT 
corrections to the LT results. We work in a framework \cite{QS}
in which the twist-four contributions can be expressed as the convolution 
of the partonic hard parts and four-parton matrix elements \cite{LQS1}. 
At the lowest order, rescattering without gluon radiation as shown in 
Fig.~\ref{fig1}(a) broadens the transverse momentum of the leading 
jet \cite{guo1} but contribute little to parton energy loss.
One can also neglect rescattering with another quark in Fig.~\ref{fig1}(b).

The dominant HT contributions to the QCD evolution of the fragmentation 
functions come from radiative processes involving rescattering 
with a gluon from another nucleon as illustrated by the central-cut diagram 
in Fig.~\ref{fig1}(c). Kinematics only allows two poles, one at each side of 
the central-cut, out of the four propagators in the diagram. This leads to 
four possible combinations each give different momentum fractions to the 
initial partons. In one case, the initial gluon has $x_2=x_L+x_D$
which is finite when $k_T\rightarrow 0$, where
\begin{eqnarray}
  x_L&=&\frac{\ell_T^2}{2p^+q^-z(1-z)} \,\, ; \,\,
  x_D=\frac{k_T^2-2\vec{k}_T\cdot \vec{\ell}_T}{2p^+q^-z} \, ,
\label{eq:xld}
\end{eqnarray}
$\ell_T$ is the transverse momentum of the radiated gluon,
$k_T$ is the initial gluon's intrinsic transverse momentum, 
and $z=\ell_q^-/q^-$ is the momentum fraction carried by the final quark.
This corresponds to gluon radiation induced by the rescattering and
is referred to as a double-hard process.
In another combination, $x_2=x_D$ which vanishes when $k_T\rightarrow 0$. 
In this case the rescattering is soft and the gluon radiation is induced
by the initial hard photon-quark scattering. Such a process is called hard-soft.
The four contributions from Fig.~\ref{fig1}(c) correspond to these two distinct 
processes and their interferences. Their sum has the form,
\begin{eqnarray}
H^{D(1)}_{\mu\nu}&\propto&
(1-e^{-ix_Lp^+y_2^-})(1-e^{-ix_Lp^+(y^- - y_1^-)}) \nonumber \\
&\times& e^{ix_Dp^+(y^-_1-y^-_2)}.
\end{eqnarray}
This clearly manifests the LPM interference pattern caused by
the destructive interferences between hard-soft and double-hard processes.
The interference pattern is dictated by the gluon's formation time,
$\tau_f \equiv 1/x_Lp^+$,
relative to the nuclear size. The two processes completely cancel each
other in the collinear limit when $\ell_T\rightarrow 0$. Diagrams involving
three-gluon vertices have exactly the same structure as Fig.~\ref{fig1}(c),
except that they have different momentum dependence and color factor in the
hard part.

We have considered all together 23 possible cut diagrams, 14 of them are 
interferences between no and double rescattering (shown as the left and 
right-cut diagrams in Fig.~\ref{fig1}(c)) which cancel some of the contributions
from central-cut diagrams. Including virtual corrections, we obtain \cite{long} 
the leading HT contribution from rescattering  processes,
\begin{eqnarray}
\frac{dW_{\mu\nu}^D}{dz_h}
&=&\sum_q \,e_q^2 \int dx H^{(0)}_{\mu\nu}(x,p,q)
\frac{2\pi\alpha_{\rm s}}{N_c}\int \frac{d\ell_T^2}{\ell_T^4}  
\int_{z_h}^1\frac{dz}{z} \nonumber \\
&\times& D_{q\rightarrow h}(z_h/z) \frac{\alpha_{\rm s}}{2\pi} C_A
\left[ \frac{1+z^2}{(1-z)_+}T^A_{qg}(x,x_L) \right. \nonumber \\
&+&\left.\delta(z-1) \Delta T^A_{qg}(x,\ell_T^2) \right] \, ,\label{eq:WD-fq2}
\end{eqnarray}
where
\begin{eqnarray}
T^A_{qg}(x,x_L)&=& \int \frac{dy^{-}}{2\pi}\, dy_1^-dy_2^-
e^{i(x+x_L)p^+y^-+ix_Tp^+(y_1^--y_2^-)} \nonumber \\
&\frac{1}{2}&\langle A | \bar{\psi}_q(0)\,
\gamma^+\, F_{\sigma}^{\ +}(y_{2}^{-})\, F^{+\sigma}(y_1^{-})\,\psi_q(y^{-})
| A\rangle \nonumber \\
&\times&(1-e^{-ix_Lp^+y_2^-}) (1-e^{-ix_Lp^+(y^--y_1^-)}) \nonumber \\
 &\times&\theta(-y_2^-)\theta(y_2^--y_1^-)\, , \nonumber \\
\label{eq:qgmatrix}
\end{eqnarray}
is quark-gluon correlation function which essentially contains four independent
four-parton matrix elements in a nucleus and 
$x_T=\langle k_T^2\rangle/2p^+q^-=x_B\langle k_T^2\rangle/Q^2$.
With the definition of the $+$ functions \cite{AP}, the term proportional to 
the $\delta$-function accounts for virtual corrections and
\begin{eqnarray}
\Delta T^A_{qg}(x,\ell_T^2) \equiv
\int_0^1 dz\frac{1}{1-z}&[&2T^A_{qg}(x,x_L)|_{z=1} \nonumber \\
&-&(1+z^2)T^A_{qg}(x,x_L)] . \label{eq:vsplit}
\end{eqnarray}
One can similarly get the contribution from the gluon fragmentation.
We will neglect the radiative corrections to processes such as Fig.~\ref{fig1}(b)
that involve rescattering with a quark in the leading log approximation, 
because they can be shown to be proportional to $1/\ell_T^2$ \cite{long} 
as compared to  $1/\ell_T^4$ in Eq.~(\ref{eq:WD-fq2}).

Summing up all the leading contributions from LT and  HT 
processes, we can effectively define the modified quark 
fragmentation function as
\begin{equation}
\frac{dW_{\mu\nu}}{dz_h}=\sum_q e_q^2\int dx f_q^A(x,\mu_I^2) 
H^{(0)}_{\mu\nu}(x,p,q)
\widetilde{D}_{q\rightarrow h}(z_h,\mu^2). \label{eq:Wtot}
\end{equation}
where for completeness $f_q^A(x,\mu_I^2)$ should also include the
HT contributions as studied by Mueller and Qiu \cite{MQ}.
The modified quark fragmentation function satisfies the following 
evolution equation
\begin{eqnarray}
  \frac{\partial \widetilde{D}_q(z_h,\mu^2)}{\partial \ln \mu^2}  &=& 
  \frac{\alpha_{\rm s}}{2\pi} \int^1_{z_h} \frac{dz}{z} 
\left[ \widetilde{\gamma}_{q\rightarrow qg}(z,x,x_L,\mu^2)
\widetilde{D}_q(z_h/z,\mu^2) \right. \nonumber \\
&+&\left.\widetilde{\gamma}_{q\rightarrow qg}(1-z,x,x_L,\mu^2) 
D_g(z_h/z,\mu^2)\right] \, ,  \label{eq:e-ap1}
\end{eqnarray}
with the modified splitting functions defined as
\begin{eqnarray}
\widetilde{\gamma}_{q\rightarrow qg}(z,x,x_L,\ell_T^2) &=&
\gamma_{q\rightarrow qg}(z)+\Delta\gamma_{q\rightarrow qg}(z,x,x_L,\ell_T^2)
\nonumber \\
& &\label{eq:e-split1}\\
\Delta\gamma_{q\rightarrow qg}(z,x,x_L,\ell_T^2)&=&
\frac{2\pi\alpha_{\rm s}C_A}
{\ell_T^2 N_c f_q^A(x,\mu_I^2)} 
\left[\frac{1+z^2}{(1-z)_+}T^A_{qg}(x,x_L) \right.\nonumber\\
&+& \left.\delta(1-z)\Delta T^A_{qg}(x,\mu^2) \right], \nonumber \\
& &
\label{eq:dsplit1}
\end{eqnarray}
where $\gamma_{q\rightarrow qg}(z)$ is the normal splitting 
functions \cite{Fields}.
We assume in the leading order that the gluon fragmentation function 
follows the normal QCD evolution equations.

Solving the above equation is equivalent to summing all leading log 
twist-four contributions.
As an approximation, one can write the
solution as,
\begin{eqnarray}
\widetilde{D}_{q\rightarrow h}(z_h,\mu^2)&\equiv& 
D_{q\rightarrow h}(z_h,\mu^2) +\Delta D_{q\rightarrow h}(z_h,\mu^2) 
\nonumber \\
\Delta D_{q\rightarrow h}(z_h,\mu^2) &=&\frac{\alpha_{\rm s}}{2\pi} 
\int_0^{\mu^2} \frac{d\ell_T^2}{\ell_T^2} 
\int_{z_h}^1 \frac{dz}{z} \nonumber \\
&\times&\left[ \Delta\gamma_{q\rightarrow qg}(z,x,x_L,\ell_T^2) 
D_{q\rightarrow h}(z_h/z,\mu^2) \right. \nonumber \\
&+& \left. \Delta\gamma_{q\rightarrow gq}(z,x,x_L,\ell_T^2)
D_{g\rightarrow h}(z_h/z,\mu^2)\right] \, , \nonumber \\
& &\label{eq:dmod}
\end{eqnarray}
where $D_{a\rightarrow h}(z_h,\mu^2)$ are the normal fragmentation functions.
Notice that there is no collinear divergence in the above integration because
of the LPM effect in $T_{qg}^A$.
Because $\Delta\gamma$ is proportional to $1/\ell_T^2$, 
$\Delta D_{q\rightarrow h}$ is suppressed by $1/\mu^2$ relative to
the LT fragmentation function $D_{q\rightarrow h}$.

To estimate the twist-four parton matrix elements, we generalize the
approach by \cite{LQS1} to include the off-diagonal matrix elements.
Assuming a Gaussian nuclear distribution in the rest frame,
$\rho(r)\sim \exp(-r^2/2R_A^2)$, $R_A=1.12 A^{1/3}$ fm, 
we express $T_{qg}^A$ in terms of single parton distributions,
\begin{eqnarray}
T^A_{qg}(x,x_L)&=&\frac{C}{x_A}\left[ f_q^A(x)(x_T+x_L)G(x_T+x_L) 
\right.\nonumber \\
&+&\left.f_q^A(x+x_L)x_TG(x_T) \right] (1-e^{-x_L^2/x_A^2}),
\label{eq:tqg2}
\end{eqnarray}
where $G(x)$ is the gluon distribution per nucleon in a nucleus, 
$x_A=1/MR_A$, and $M$ is the nucleon's mass. The off-diagonal terms involves 
transferring momentum $x_L$ between different nucleons inside a nucleus
and thus should be suppressed for large nuclear size or large momentum 
fraction $x_L$. Notice that $\tau_f=1/x_Lp^+$ is the gluon's formation time. 
Thus, $x_L/x_A=L_A/\tau_f$ with $L_A=R_AM/p^+$ being the nuclear size in our
chosen frame.

\begin{figure}
\centerline{\psfig{figure=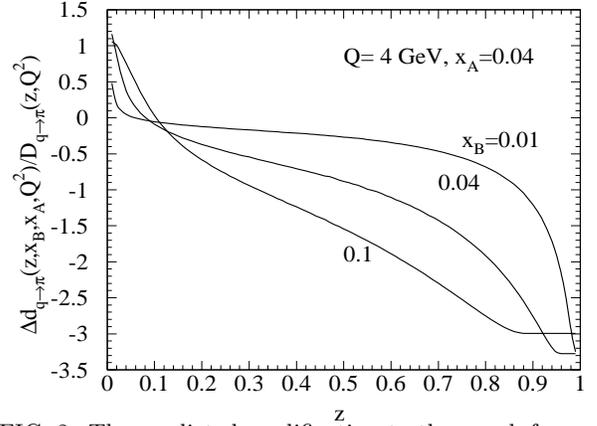,width=3.0in,height=2.2in}}
\caption{The predicted modification to the quark fragmentation 
functions for three different values of initial quark energy 
$q^-=Q^2/2p^+x_B$. $x_A=0.04$ corresponds to $A\approx 200$}
\label{fig2}
\end{figure}

Using the above approximation in Eq.~(\ref{eq:dmod}) and replacing the momentum
fraction $x_L$ in the parton distributions by its average value
$\langle x_L\rangle \sim x_A$, we have
\begin{eqnarray}
\Delta D_{q\rightarrow h}(z_h,Q^2)&\approx&
\frac{C_A\alpha_{\rm s}^2}{N_c}\frac{\widetilde{C}}{Q^2x_A^2}
\Delta d_{q\rightarrow h}(z_h,x_B,x_A,Q^2), \nonumber \\
\widetilde{C}&=&C[x_TG(x_T)+(x_T+x_A)G(x_T+x_A)], \nonumber \\
& &\label{eq:modf}
\end{eqnarray}
where we choose the factorization scale $\mu^2=Q^2$. Shown in Fig.~2 are the 
numerical results of $\Delta d(z_h,x_B,x_A,Q^2)$ for three different values 
of $x_B$. The parameterization in Ref.~\cite{bkk} of the normal fragmentation
functions is used in our calculation. As we see from the numerical results, 
the modification to the shape of
fragmentation function increases for larger values of $x_B$ corresponding to
smaller quark energy $q^-$ at fixed $Q^2$. The modification also increases
for smaller values of $x_A$ corresponding to larger nuclear size. As shown
in Eq.~(\ref{eq:modf}), the magnitude of the modification depends quadratically
on the nuclear size. This is because the LPM effect modifies the transverse
momentum spectra of the radiated gluon such that the phase space for the
momentum integral is limited. That limited phase space is proportional to
the nuclear size. Together with the linear dependence of the parton correlation
function, this leads to non-linear dependence of the total energy loss on the 
nuclear size.

We can define the quark energy loss as the momentum carried 
by the emitted gluons. From Eq.~(\ref{eq:dmod}), we have,
\begin{eqnarray}
\langle\Delta z_g\rangle&=& \int_0^{Q^2}\frac{d\ell_T^2}{\ell_T^2} 
\int_0^1 dz \frac{\alpha_{\rm s}}{2\pi}
 z\,\Delta\gamma_{q\rightarrow gq}(z,x,x_L,\ell_T^2) \nonumber \\
&\approx & \frac{C_A\alpha_{\rm s}^2}{N_c}\frac{\widetilde{C}}{Q^2}
\frac{x_B}{x_A^2} 6\ln(\frac{1}{2x_B})\; , \;\;\; (x_A\ll x_B\ll 1).
\end{eqnarray}
According to Ref.~\cite{guo2,GQZ}, $\widetilde{C}$ can be related to the 
transverse momentum broadening of Drell-Yan dilepton in $pA$ collisions,
\begin{equation}
\Delta\langle k_T^2\rangle_{\rm DY} \approx\frac{2\pi\alpha_{\rm s}}{N_c}\frac{\widetilde{C}}{2x_A}
\approx0.022 A^{1/3}\; {\rm GeV}^2.
\end{equation}
In the rest frame of the nucleus, the total energy loss is
\begin{equation}
\Delta E=q_0 \langle\Delta z_g\rangle \approx 0.35\; \alpha_{\rm s} A^{2/3} 
\ln\frac{1}{2x_B} \;\; {\rm GeV} \; ,
\end{equation}
which depends quadratically on the nuclear size. For $x_B=0.1$, $\alpha_{\rm s}=0.3$
and $A=200$, $\Delta E\approx 5.8$ GeV which is consistent with the 
estimate in Ref~\cite{BDMPS}.

In summary, we have derived for the first time the evolution equations for the
nuclear modified quark fragmentation functions due to multiple scattering and
induced radiation. 
We can generalize our results to the case of hot QCD matter. In our approximation,
Eq.~(\ref{eq:tqg2}), the quark distribution which represents the initial quark
production probability does not enter into the modified evolution equations.
What remains is the gluon matrix element. In the case of a hot QCD medium, the
coordinates of the two gluon fields are not restricted to the nucleon size 
due to deconfinement in a nucleus. Rather, the final results will be 
sensitive to the correlation length of the hot medium \cite{screen}. If there
is a dramatic change of the correlation length during the QCD phase transition,
one should also see a big change in the modification of the parton
fragmentation functions.

\section*{Acknowledgements}
We would like to thank J. Jalilian-Marian, J. Qiu and G. Sterman for helpful 
discussions. This work was supported by DOE under Contract No. DE-AC03-76SF00098 
and DE-FG-02-96ER40989 and in part by NSFC under project 19928511.

\end{multicols}
\end{document}